\documentclass{article}
\usepackage{spconf,amsmath,graphicx}
\usepackage{multirow}
\usepackage{xcolor}
\usepackage{tikz}
\usetikzlibrary{positioning}
\usepackage[pscoord]{eso-pic}
\usepackage{framed}
\usepackage{booktabs}
\usepackage{caption}
\usepackage{subcaption}
\usepackage{url}
\newcommand{\accessedDate}{Oct.~2020}
\newcommand{\urlfootnote}[1]{\footnote{\url{#1} Accessed \accessedDate}}

\newcommand{\placetextbox}[3]{
  \AddToShipoutPictureFG*{
    \put(\LenToUnit{#1\paperwidth},\LenToUnit{#2\paperheight}){\vtop{{\null}
        \framebox[\textwidth]{\parbox{\dimexpr\textwidth-2\fboxsep-2\fboxrule}{\footnotesize{#3}}}  
    
    }}%
  }%
}%

\title{Effect of Language Proficiency on Subjective Evaluation of Noise Suppression Algorithms}

\name{Babak Naderi$^1$, Gabriel Mittag$^1$, Rafael Zequeira Jim\a'enez$^1$, Sebastian Möller$^{1,2}$}
\address{$^1$ Quality and Usability Lab, Technische Universität Berlin, Berlin, Germany\\
$^2$ Language Technology, Deutsches Forschungszentrum für Künstliche Intelligenz (DFKI), Berlin, Germany}

\begin{document}
%
\maketitle
\begin{abstract}
Speech communication systems based on Voice-over-IP technology are frequently used by native as well as non-native speakers of a target language, e.g. in international phone calls or telemeetings. Frequently, such calls also occur in a noisy environment, making noise suppression modules necessary to increase perceived quality of experience. Whereas standard tests for assessing perceived quality make use of native listeners, we assume that noise-reduced speech and residual noise may affect native and non-native listeners of a target language in different ways. To test this assumption, we report results of two subjective tests conducted with English and German native listeners who judge the quality of speech samples recorded by native English, German, and Mandarin speakers, which are degraded with different background noise levels and noise suppression effects. The experiments were conducted following the standardized ITU-T Rec. P.835 approach, however implemented in a crowdsourcing setting according to ITU-T Rec. P.808. Our results show a significant influence of language on speech signal ratings and, consequently, on the overall perceived quality in specific conditions. 

\end{abstract}
\section{Introduction}
\placetextbox{0.09}{0.07}{©2020 IEEE. Personal use of this material is permitted. Permission from IEEE must be obtained for all other uses, in any current or future media, including reprinting/republishing this material for advertising or promotional purposes, creating new collective works, for resale or redistribution to servers or lists, or reuse of any copyrighted component of this work in other works.}

Speech calls, especially when performed on mobile phones, are commonly carried out in noisy environments. To ensure an acceptable quality for the communication partners, noise suppression algorithms are commonly implemented into the end-user device or the network. To evaluate the effectiveness of such algorithms, different types of perceptual degradations which essentially stem from noise miss-estimations have to be considered: 1) residual noise may result in perceived noisiness; 2) additive artefacts (e.g. musical tones may cause perceptual discontinuity; 3) subtraction of speech signals may result in degraded intelligibility and discontinuity. In order to address the diversity of perceptual phenomena and their impact on overall perceived quality, noise suppression algorithms are commonly evaluated in laboratory-based listening tests according to ITU-T Rec. P.835, where participants subsequently and separately evaluate the quality of the speech signal, the background noise, and the overall quality of degraded and enhanced speech stimuli. 
According to ITU-T Rec. P.835, all participants of such a listening test should be native speakers of the target language used in the test. However, this setting might not reflect realistic communication use cases, in which frequently, non-native listeners of a language are involved. It can be assumed that noisy, as well as noise-reduced speech showing the perceptual artefacts described above, will affect native and non-native listeners in different ways, and to different degrees. This effect, however, has to our knowledge never been analyzed in a scientific way. In addition, laboratory tests as the one described in ITU-T Rec. P.835 are more and more replaced by crowdsourcing-based online tests, carried out by paid participants (crowdworkers) who may not be native in the target language. Thus, it is important to check whether the listeners language proficiency shows an impact on the ratings in order to validate such test settings.

In the following, we report two listening tests in which German, English and Mandarin noisy and noise-reduced speech files were judged by native as well as non-native listeners of the target language. Section 2 reviews related work; Section 3 describes the experimental set-up; Section 4 analyzes the results obtained both with respect to the effects of noise suppression on speech, background noise and overall quality, and with respect to the validity of crowdsourcing experiments for this purpose. Finally, Section 5 discusses the findings and proposes steps for future work.

\section{Related works}
%

Multiple studies have analyzed different aspects of conducting an audio listening test with users of different nationalities. For instance, work in~\cite{ITU-P.Sup23} presents multiple P.800 listening tests conducted with listeners from different countries that assessed the quality of speech files in their native language. Authors found that Japanese listeners provided lower quality scores per condition than French, English, German and Norwegian listeners. 
However, no analysis was made regarding the language's influence on the quality scores.

Authors in \cite{Blaskova2008} present the results of a P.800 listening test conducted with users of a mother tongue different from the speech samples to be assessed. Listeners were either native Czech, Slovak, or Italian speakers. They assessed the quality of an English speech dataset containing different codecs and two noise conditions. Authors found that listeners with insufficient English knowledge rated the speech quality systematically lower than the participants with an advanced level. Authors believe this was due to the listeners' inability to understand what was said in the stimulus. Contrary to our work, the listening test was conducted according to the Rec. P.800, whereas our study followed Rec P.835. Also, no insight was given on how listeners from different countries perceived the quality of noisy speech. 

Furthermore, \cite{Schinkel-Bielefeld2017} investigates whether foreign language influences ratings and listening times in a subjective evaluation of audio with intermediate impairments (a.k.a MUSHRA \cite{ITU-R-BS.1534-32014}). The authors conducted a study containing native German and Mandarin Chinese speaking listeners and items of these two languages. They found that for high audio quality items, non-native listeners executed more comparisons and needed 20\% more time to conduct the listening test. Additionally, no overall difference was found between the ratings of the native and non-native listeners. Unlike our work, the quality of items in a MUSHRA test is higher, and intelligibility is not an issue since listeners focus on small audio sections without semantic meaning. In contrast, we investigate how listeners of different mother tongues perceive noise-degraded speech.

\section{Method}
We created a dataset containing the so called \textit{Reference Conditions} for English, German and Mandarin target languages. As source speech signals, we used the clean fullband files of ITU-T P.501 Annex C \cite{ITU-P501} which includes four different source files (two male, two female speakers) per language. 

\subsection{Reference Conditions}
In every P.835 experiment, twelve reference conditions should be used to anchor the ratings of the test participants. These conditions independently vary the level of signal and background noise distortions to result in ratings of the entire range of values on the different scales \cite{ITU-P835}. 

Given the source speech signals, we applied the tools provided in \cite{S4_160397} to process each file with the 12 reference conditions proposed in ETSI\,TS\,103\,281 \cite{ETSI_TS_103_281}. As preprocessing step, the source signals are filtered with a 20\,kHz lowpass filter and then normalized to an active speech level of -26\,dBov, both functions are taken from ITU-T G.191 \cite{ITU-G191}. The twelve conditions are presented in Table \ref{tab:rc}, where the first condition represents the clean signal, which is the best anchor condition for all three scales. The conditions 2--5 are processed with additive car background noise at different SNR levels, and therefore affect the background noise but not the signal quality.
Conditions 6--9 are processed with a \textit{spectral subtraction distortion} that was proposed in \cite{AH-11-029}. It is based on the Wiener filter and leads to speech signal degradation similar to signal distortions evoked by noise suppression algorithms. This type of distortion only affects signal quality while no background noise is added, where a lower ``NS level'' corresponds to a higher distortion level.
Conditions 10--12 are processed with both types of distortions and represent a mix of background noise and signal distortion. 

\begin{table}[ht]
    \caption{Reference conditions for fullband subjective evaluation of noise reduction according to ETSI\,TS\,103\,281 \cite{ETSI_TS_103_281}.}
    \label{tab:rc} 
    \small
    \begin{center}
        \begin{tabular}{ l c c c}
        \toprule
        \textbf{Con} & \textbf{Speech Distortion}&	\textbf{Noise Type}&	\textbf{SNR (A)} \\ 
        \midrule
        i01 & - &  - & -\\
        i02 & -  & Car 130\,kmh &  0 dB\\
        i03 & -  &  Car 130\,kmh & 12 dB\\
        i04 & - &  Car 130\,kmh & 24 dB\\
        i05 & - &  Car 130\,kmh  &36 dB\\
        \midrule
        i06 & NS Level 1 &-& -\\
        i07 & NS Level 2 &-& -\\
        i08 & NS Level 3 &-& -\\
        i09 & NS Level 4 &-& -\\
        \midrule
        i10 & NS Level 3 &  Car 130\,kmh  & 24 dB\\
        i11 & NS Level 2 &  Car 130\,kmh &  12 dB\\
        i12 & NS Level 1 &  Car 130\,kmh &  0 dB\\
        \bottomrule
        \end{tabular}
    \end{center}
    \vspace{-0.5cm}
\end{table}

\subsection{Subjective test}
We conducted two subjective crowdsourcing tests using the open-source P.808 Toolkit\cite{naderi2020open}, which was recently extended to provide tests in line with ITU-T Rec. P.835\urlfootnote {https://github.com/microsoft/P.808}.
In the P.835 test procedure, participants are instructed to rate the \textit{background noise}, the \textit{speech signal}, and the \textit{overall quality} of a speech sample on corresponding scales in one trial. Before each rating, the participant should listen to the speech sample again and is instructed to only attend to that specific aspect.
The rating section of the crowdsourcing task contained 11 trials including one trapping question and one gold standard question. Besides that, the crowdsourcing task contained a qualification section (including the hearing test), a setup section (including environment suitability \cite{naderi2020env} and headset usage tests) and a training section (including anchoring speech samples) as specified in the ITU-T Rec. P.808 \cite{ITU-P808}.

In Exp1, we used the Amazon Mechanical Turk (AMT) crowdsourcing platform to recruit native English speaking participants from the US. In Exp2, we recruited native German speakers using the ClickWorker crowdsourcing platform.
In both experiments, participants rated the above-mentioned dataset. 
We performed the data screening process according to the ITU-T Rec. P.808 and removed submissions that failed in one of the trapping and gold standard questions, hearing, environment and hardware usage test, or with specific voting pattern. Following these criteria, 86.7\% of submissions in Exp1 and 74.5\% in Exp2 are considered reliable. On average we collected 73 and 62 valid votes per test conditions (i.e. per degradation in each language) in Exp1, and Exp2 respectively. 

\section{Results}
\begin{figure*}[ht]
    \begin{subfigure}[b]{0.3\textwidth}
        \includegraphics[width=1\textwidth]{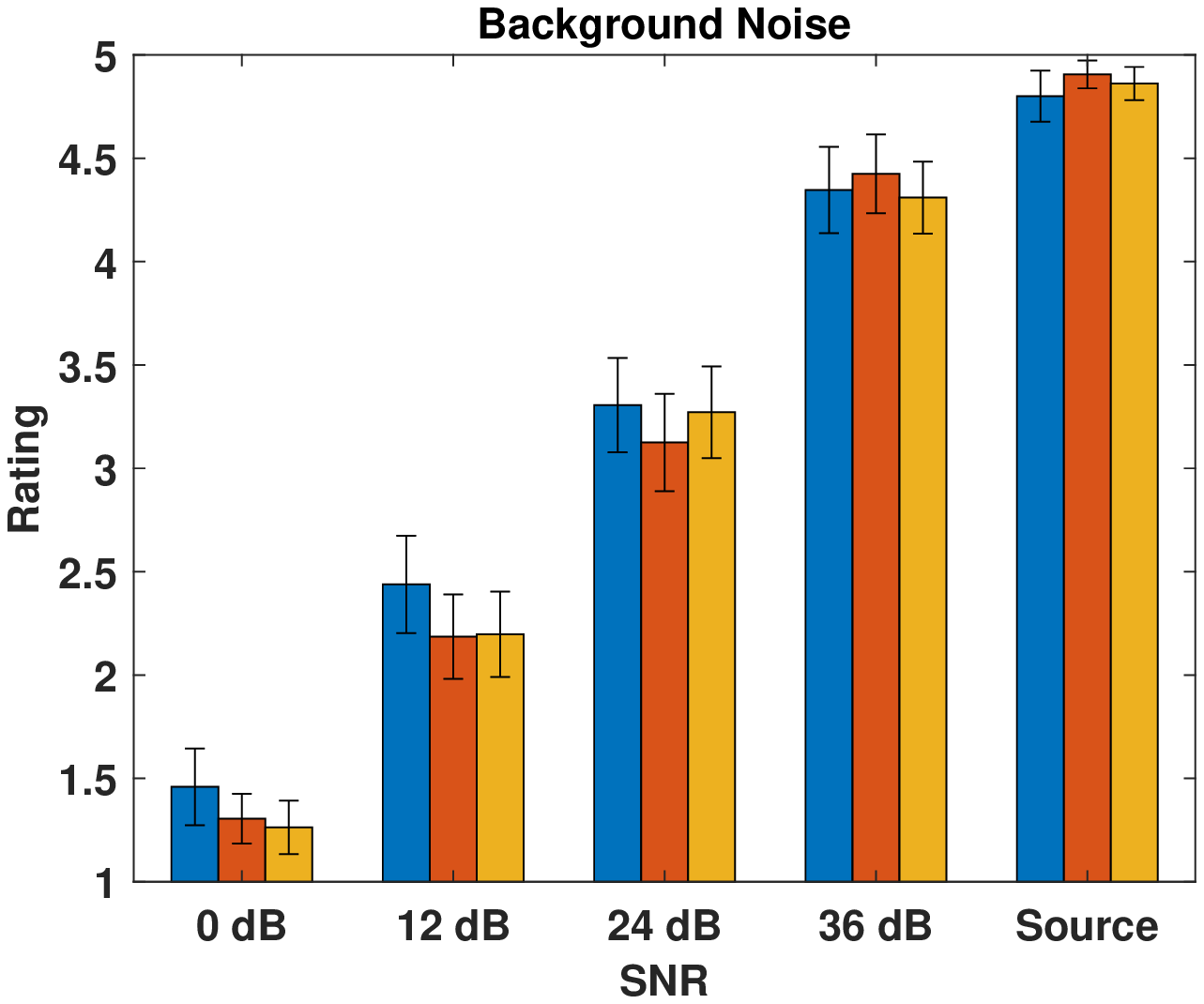}
        \caption{}
    \end{subfigure}    
     ~
    \begin{subfigure}[b]{0.3\textwidth}
        \includegraphics[width=1\textwidth]{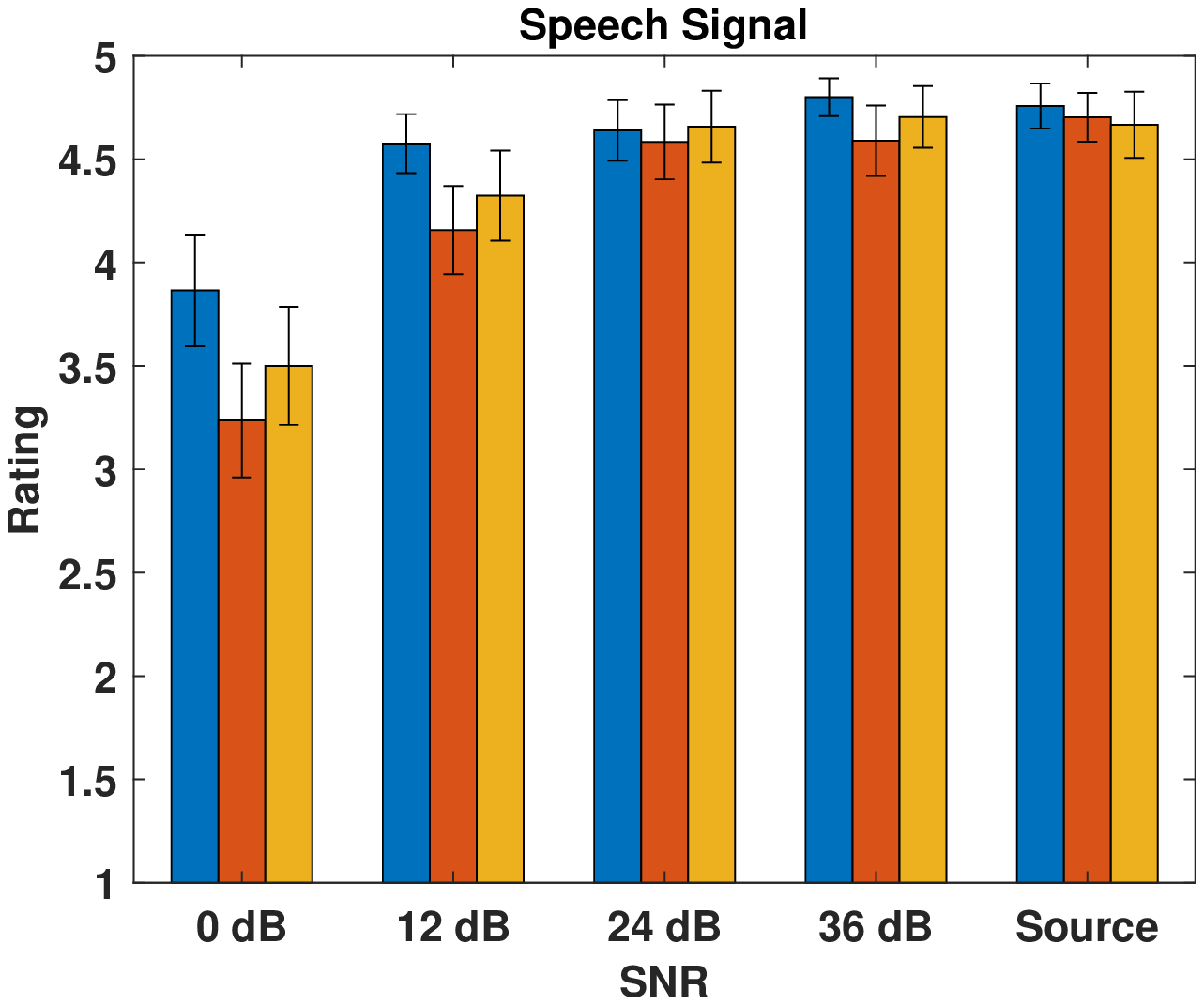}
        \caption{}
    \end{subfigure}   
     ~
    \begin{subfigure}[b]{0.3\textwidth}
        \includegraphics[width=1\textwidth]{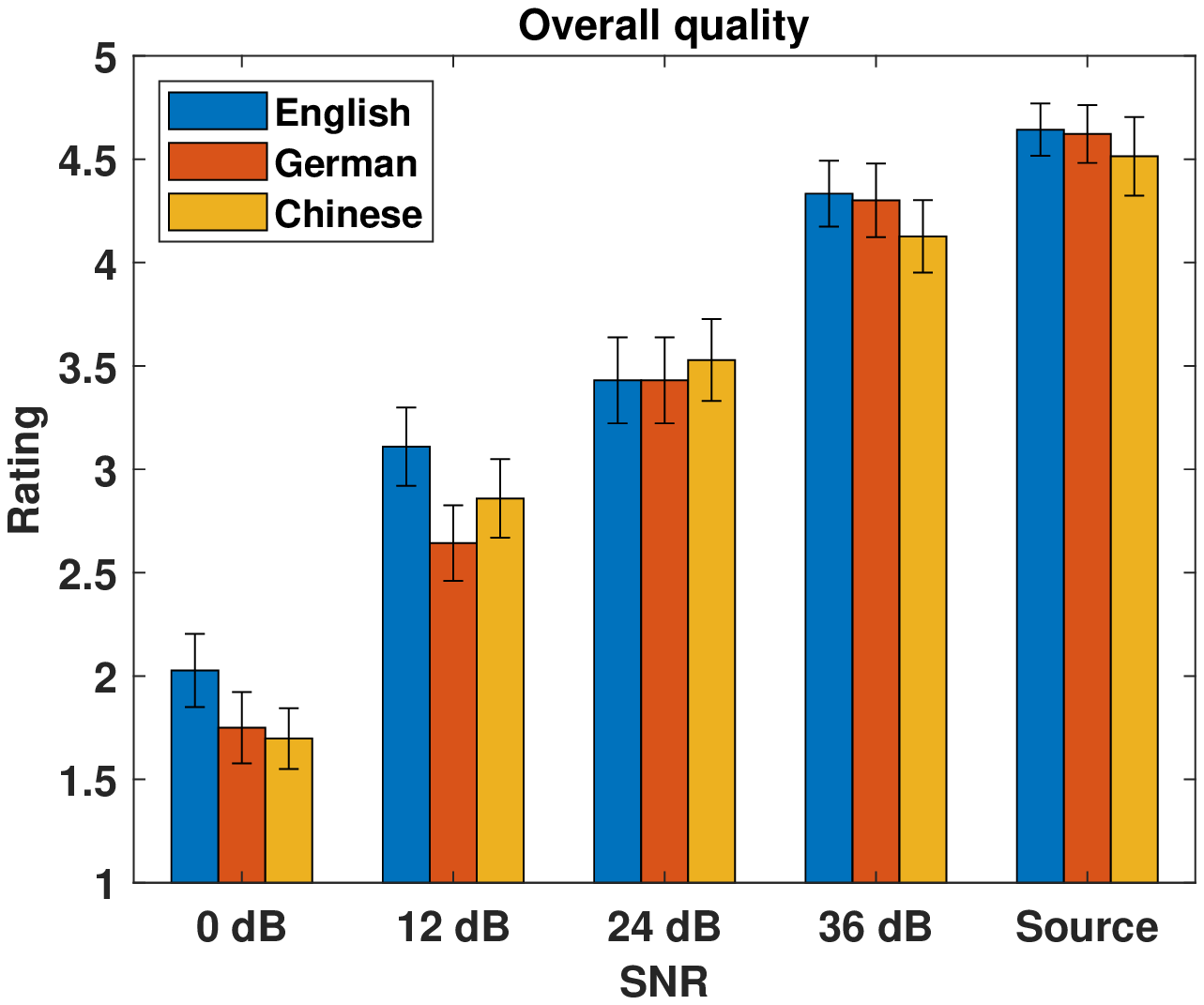}
        \caption{}
    \end{subfigure}  
     \\
    \begin{subfigure}[b]{0.3\textwidth}
        \includegraphics[width=1\textwidth]{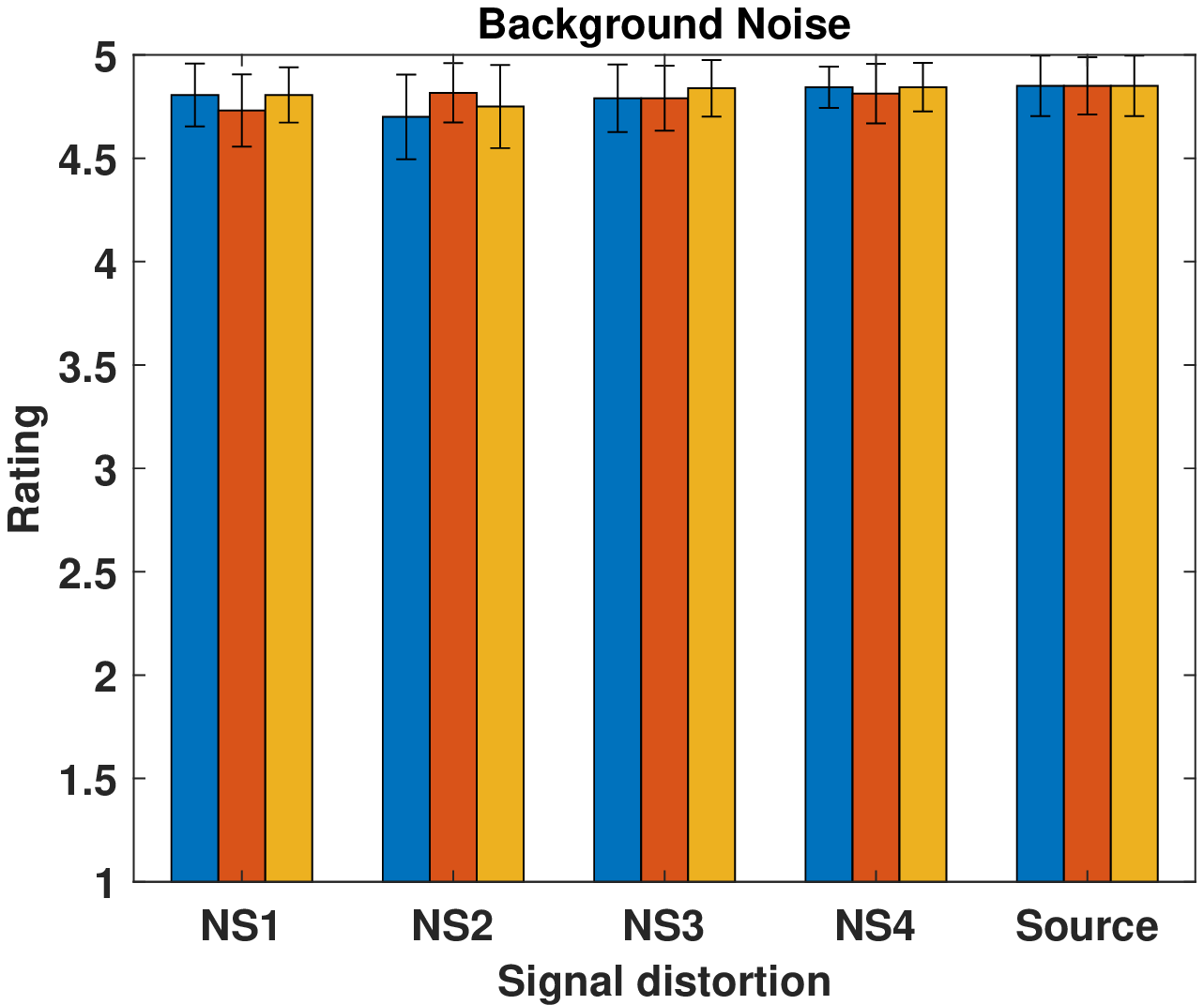}
        \caption{}
    \end{subfigure}    
     ~
    \begin{subfigure}[b]{0.3\textwidth}
        \includegraphics[width=1\textwidth]{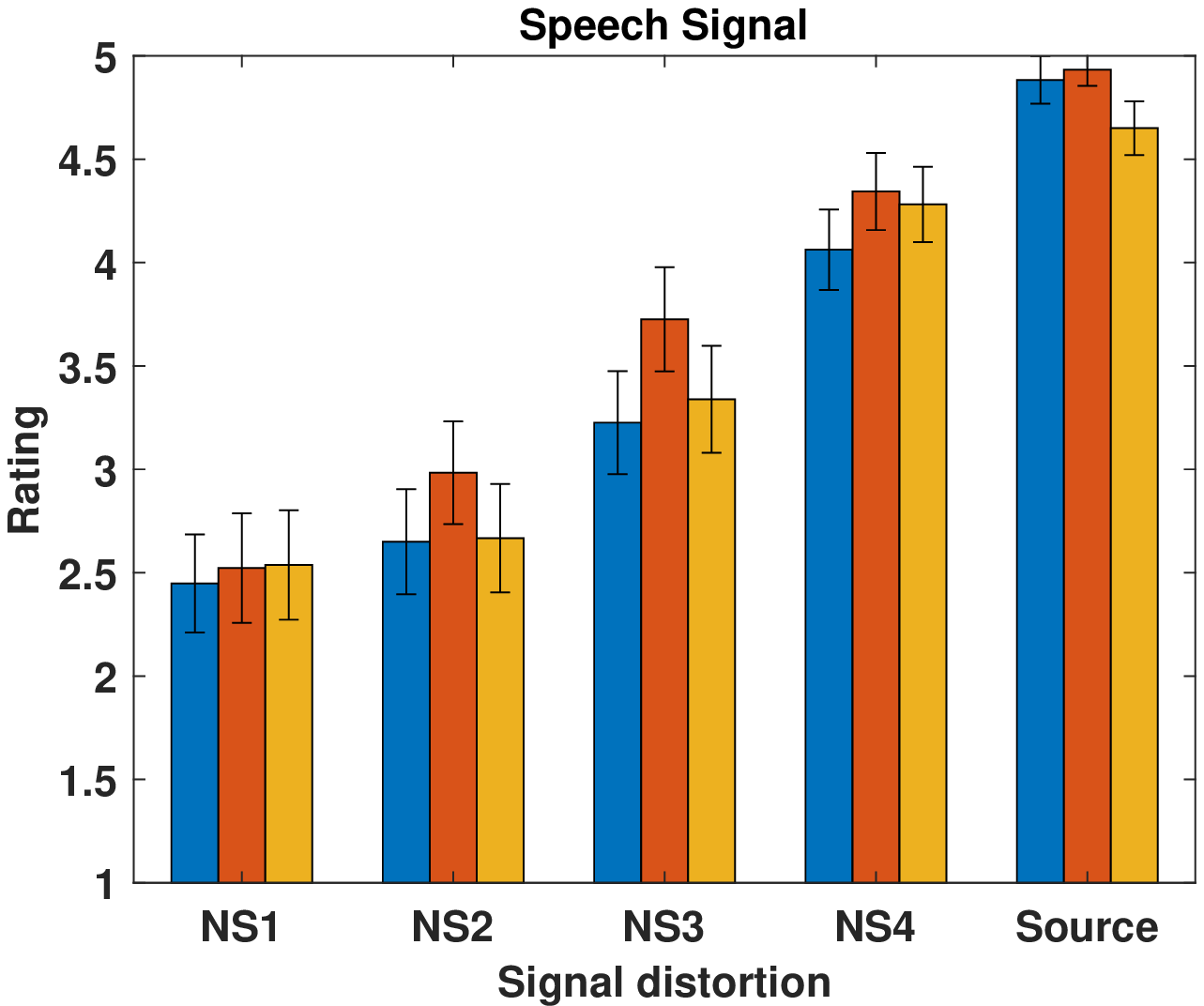}
        \caption{}
    \end{subfigure}   
     ~
    \begin{subfigure}[b]{0.3\textwidth}
        \includegraphics[width=1\textwidth]{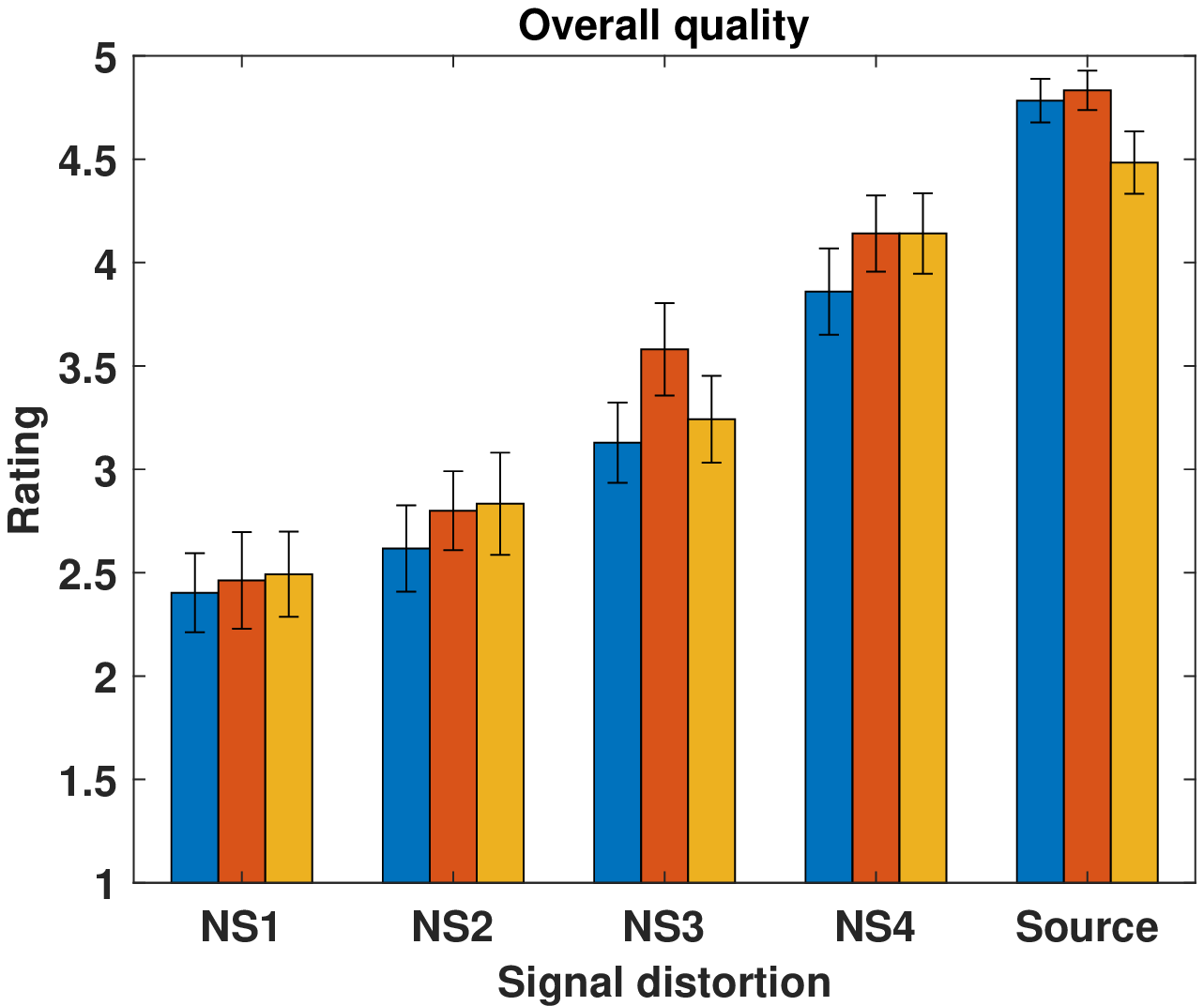}
        \caption{}
    \end{subfigure}  
    \caption{Auditory results of the selected reference conditions: (a-c) Exp1 (native English participants), background noise varies, signals not distorted i.e. i02-5 and i01, respectively. 
    (d-f) Exp2 (native German participants), no background noise, signal distortion varies i.e. i06-9 and i01, respectively.}
    \label{fig:results:all}
\end{figure*}

We calculated the Mean Opinion Score (MOS) for each of the signal, background, and overall quality rating per degradation and per language in Exp1 (native English participants) and Exp2 (native German participants).
Overall, ratings on the target language of which the participant was a native speaker were highly correlated to ratings on other languages for all three scales (on average PCC=0.98 in Exp1 and PCC=0.99 in Exp2). The smallest correlations were observed for ratings on the signal scale, between the German and the English language in both experiments (PCC=0.97 in Exp1, and PCC=0.98 in Exp2).
Figure \ref{fig:results:all} (a-c) illustrates the MOS values over the three scales for conditions in which the SNR varied but no signal degradation was employed in Exp1. 
Results show that in presence of strong background noise, participants perceived the signal and overall quality lower when they rated a dataset in other languages than their native one. 

In Exp2, we organized the stimuli sets for crowdsourcing sessions in a way that one worker rated the same degradation condition in three languages. As a result, we could conduct a two-way Mixed Analysis of Variance (ANOVA) to compare ratings on each scale using target language as a within-subject factor and degradation condition as a between-subject factor.

For the signal scale, the ANOVA yielded a significant interaction effect of language and degradation, F\,(20.86, 1353.86)$\,=\,$2.37, p$\,<\,$.001, $\eta_p^2$\,=\,.04, significant simple main effects of the degradation for all languages (c.f. Table \ref{tab:stat}), as well as significant simple main effects of the language in 6 degradation conditions namely i01-2, i07-10\footnote{For condition i02 the effect was between EN an ZH, for the rest the difference was between DE and at least one of the others.}. Multiple comparisons between different degradations revealed that in 8 cases the EN dataset, and in 5 cases the ZH dataset led to different conclusions\footnote{Conclusion here means either two degradation conditions are significantly different or not.} than the DE dataset, which was the native language of the test participants (i.e. 12\% and 8\% disagreements, respectively).

For the background noise scale, the ANOVA only yielded a significant main effect of degradation, F\,(11, 714)$\,=\,$340.77, p$\,<\,$.001, $\eta_p^2$\,=\,.84. Results of post-hoc tests showed statistically significant differences between degradation conditions when different SNR values applied.

For the overall quality rating, the ANOVA yielded a significant interaction effect of language and degradation, F\,(21.27, 1380.53)$\,=\,$3.06, p$\,<\,$.001, $\eta_p^2$\,=\,.05, significant simple main effects of the degradation, for all three languages (c.f. Table \ref{tab:stat}), as well as significant simple main effects of the language in 5 degradation conditions namely i01-3, i08-9\footnote{For conditions i02-3 the significant effects were between EN an ZH.}. Multiple comparisons between different degradations revealed that in 6 cases the EN dataset, and in 8 cases the ZH dataset lead to different conclusions than the DE dataset.

Figure \ref{fig:results:all} (d-f) illustrates the MOS values over the three scales for conditions in which signal degradation varied without a background noise in Exp2. 

\begin{table}[t]
    \caption{ANOVA statistics of simple main effect of degradation for the Signal and Overall quality scales.}
    \label{tab:stat} 
    \small
    \begin{center}
        \begin{tabular}{ l c  c c c  c  }
        \toprule
        \textbf{Scale} &	\textbf{Language}&$df_a, df_d $ &	\textbf{$F$}&	\textbf{$P$} & \textbf{ $\eta_p^2$} \\ 
        \midrule
        \multirow{3}{*}{Signal} & DE & 11,714& 63.979 &	$<$.001 &	0.496\\ 
         & EN & 11,714  & 75.256 &	$<$.001 &	0.537\\
         & ZH & 11,714  & 54.011 &	$<$.001 &	0.454\\
          \midrule
        \multirow{3}{*}{Overall} & DE& 11,714 & 117.294 &	$<$.001 &	0.641\\ 
         & EN  & 11,714 & 116.102 &	$<$.001 &	0.644\\
         & ZH & 11,714  & 94.845 &	$<$.001 &	0.594\\
        \bottomrule
        \end{tabular}
    \end{center}
    \vspace{-0.5cm}
\end{table}

\begin{table}[hbt!]
    \caption{Effects of rater's language proficiency (RLP) and degradation cond. on different scales. \textit{Sig. Con.} list of conditions with sig. difference between two participant groups.}
    \label{tab:lmem} 
    \small
    \begin{center}
    \resizebox{\columnwidth}{!}{%
        \begin{tabular}{ c l  l  l  l r l}
        \toprule
        \textbf{Target} &	\textbf{Scale}&	\textbf{Effect }&$df_a, df_d $ &	\textbf{$F$}&	\textbf{$P$} & \textbf{ Sig. Cond.} \\ 
        \textbf{Lang.} & & & & & & \\
        \midrule
        \multirow{6}{*}{EN} & Signal & RLP& 1, 129.55& 4.57 &	.034 &\\ 
        & Signal & Cond.& 11, 1516.45&	197.32& $<$.001& \\ 
        & Signal & RLP x Cond.& 11, 1516.45&	3.88& $<$.001& i06-8,10-1 \\
        & Backg. & Cond.& 11, 1515.43&	564.22& $<$.001& \\ 
        & Overall & Cond.& 11, 1496.15&	324.63& $<$.001& \\
        & Overall & RLP x Cond.& 11, 1496.15&	5.96& $<$.001 & i04,i06-9\\ 
          \midrule
        \multirow{4}{*}{DE} & Signal & Condition& 11, 1516.5& 151.29 & $<$.001 &\\ 
        & Signal & RLP x Cond.& 11, 1516.5&	2.96& $<$.001&  i02-3,6\\
        & Backg. & Cond.& 11, 1521.62&	548.29& $<$.001& \\
        & Overall & Cond.& 11, 1501.09&	292.51& $<$.001& \\
        & Overall & RLP x Cond.& 11, 1501.09&	3.51& $<$.001 &i01,3-4\\ 
        \bottomrule
        \end{tabular}
        }
    \end{center}
    \vspace{-0.5cm}
\end{table}

\subsection{Comparison between Experiments}

We also combined the ratings of Exp1 and Exp2 and compared them with respect to the target languages of the rated speech files. We assume that English listeners (from the US AMT population) do not have any knowledge of German\footnote{Our attempts on finding German speaking participants in AMT failed multiple times.}, whereas we anticipate that German listeners (from the German Clickworker platform) do know some English (as the test was conducted in English).

For each the English and the German speech data sets of Exp1 and Exp2, we fitted three linear mixed-effects models (LMEMs) with random intercepts and with rater's expected language proficiency (RLP) and degradation condition as fixed factors and participant as random factor (i.e. one model per scale). Significant effects are reported in Table \ref{tab:lmem} followed by conditions in which within that condition significant difference between two participant groups were observed (revealed by pairwise contrast using the Holm adjustment method). Besides two cases, in the rest of the cases significant differences exist between two groups of participants, the native participants perceived the quality to be higher than their non-native counterparts.

\section{Discussion and Conclusion}
Despite the high correlations between different target languages on all scales, we observed significant differences between languages. 
In the repeated-measure design, participants rated the signal and overall quality higher when they were native of the target language, in case only the signal is degraded (Exp2) or in presence of strong  background noise (Exp1).
In case of a clean signal or strong background noise, the ratings were different when the participant had no knowledge of the target language compared to the case when they did have some knowledge of it.
In between-group design, comparing ratings of a native group and participants with some knowledge about the target language, the signal was rated higher by natives when no or moderate background noise was present. In case of no background noise, the overall quality was also rated higher.
Similarly, native participants rated the signal in the presence of background noise higher than the group with no knowledge about the target language, but lower when only the signal is strongly distorted.
Natives also rated overall quality higher in case of clean signals or signals with only moderate background noise. 

The results show that background noise ratings are not affected by the participants' proficiency in the target language. However, speech signal and consequently overall quality (without dominating background noise) were perceived to have a higher quality by native participants.
The results regarding the impact of target language proficiency should be interpreted with care, as not formal language proficiency test was performed. As a real-life noise suppression algorithm may include both signal distortions and residual background noise, we recommend that subjective P.835 tests should be carried out with a representative sample of the service's target population.


\bibliographystyle{IEEEbib}
\bibliography{strings,refs}

\end{document}